# Tandem-ABALONE™ Photosensors with 2 × 2π Acceptance for Neutrino Astronomy

Ivan Ferenc Šegedin[1], Marija Šegedin Ferenc[1], Daniel Ferenc[1,2]

(1) PHOTONLAB, INC., Davis, CA, USA

(2) Department of Physics and Astronomy, University of California, Davis, CA, USA

**ABSTRACT**

The primary goal of the novel photosensor technology called ABALONE™ (U.S. Pat. 9,064,678) has been to enable the next-generation of astroparticle physics experiments to open new windows on the universe with large-area detectors of unprecedented performance, radio-purity, robustness, and integration flexibility. This foundational technology provides the means for modern, scalable and cost-effective production. The wide range of possible application-specific detector configurations can also make a difference in homeland security and medical imaging. Thus far, we have reduced to practice two such detector configurations: (i) the composite sandwich Panel assembly that hosts matrices of closely packed ABALONE units (U.S. Pat. 10,823,861), and, quite differently, the Tandem Detector Modules that host two back-to-back oriented ABALONE photosensors. This article for the first time presents the latter configuration, specifically designed to provide 2 × 2π angular acceptance for neutrino astronomy and other large-area radiation detectors.

## 1. Introduction

The primary goal of our foundational photosensor technology called ABALONE™ [1] has been to enable the next-generation of astroparticle physics experiments to open new windows on the universe with large-area detectors of unprecedented performance, radio-purity, robustness, and cost-effectiveness. The same platform technology can also make a significant difference in other application areas, primarily in large-area radiation detection for homeland security, and medical imaging.

The ABALONE™ Photosensor Technology is a modern, scalable technology. The exclusive use of fused silica or glass, and only thin films of metals allows the application of modern and proven manufacturing processes, i.e. low-cost high-volume production, while achieving unprecedented performance that includes the following [2-4]:

o Signal concentration of factor ~$10^4$ from the photocathode area to the photodiode readout area.
o High radio-purity, thanks to the chemical composition of the photosensor body, dominated only by fused silica with the addition of thin films of ultrapure materials.
o Sensitivity to UV light.
o Flexible dynamic range with tunable gain from $10^3$ up to $10^9$, without preamplifiers.
o Single-photon sensitivity and resolution.
o Sub-nanosecond timing resolution.
o Low positive ion afterpulsing rate of only $5 \times 10^{-3}$ ions/photoelectron.
o Immunity to accidental daylight exposure.
o Resistance to shock, compression, and vibration.
o Single-level high voltage of practically negligible current.

By design, the foundational ABALONE Photosensor Technology provides a wide range of possible application-specific detector configurations. Thus far, we have reduced to practice two such detector configurations: (i) the composite sandwich Panel assembly that hosts matrices of closely packed ABALONE units [5], and, quite differently, the Tandem Detector Modules that host two back-to-back oriented ABALONE photosensors [1]. In all cases, an optimal electron lens for photoelectron focusing – within the vacuum enclosure of each integrated ABALONE unit – is created by collective action of its own components, in conjunction with the entirety of the integration structure, strategically arranged behind the ABALONE units (notably, outside the vacuum enclosure).

In composite Panels, the electric field flows from each ABALONE unit in the matrix to all of its neighbors through the dielectric Panel core [5]. The (only) two electric potentials - high-voltage and ground - are universal for the entire panel, and they are distributed to each ABALONE member of the matrix through conductive surfaces in two printed circuit boards that sandwich the dielectric core in between them. This rigid, composite glass-fiber structure also provides mechanical support for the entire Panel assembly. In contrast, our Tandem Module forces the electric field between the two opposing ABALONE units to 'return' in a closed loop [1], as discussed below.

This article is organized as follows. After briefly reviewing the rationale behind the ABALONE Technology and the basic operational principle, we will focus on the design of the Tandem Detector Module, and on the experimental verification of its electron optics.

## 2. The Implications of the Foundational ABALONE Photosensor Technology for the Tandem Concept

The performance of the ABALONE technology [2-4], together with its suitability for (cost-effective) high-volume production [1], can bring the much-needed game change in the broad field of large-area radiation detection. To this day, this field relies on the expensive, essentially pre-World War II Photomultiplier Tube (PMT) technology [6], while its only modern alternative thus far, the millimeter-sized all-silicon Geiger-Mode-Avalanche Photo Diodes (G-APDs, also called SiPMs), remain too expensive for most large-area applications ($10 M/m$^2$). In addition to their high production cost, PMTs are fragile and require complex implementation. Typical large-area PMTs comprise more than 150 distinct components, which necessitate more than 300 spot welds. The chemical composition of these components in ordinary PMTs results in an intolerable level of radioactive background for detectors of rare particles and rare decays.

Our extensive studies yield an important conclusion: the reason why the PMT manufacture could never have evolved into a modern and cost-effective vacuum technology lies precisely in the combination of materials that the PMTs are made of. The same holds true for some more recent variations of the same technology, including Microchannel Plate PMTs (MCP-PMTs) [7-9], and Hybrid Photo Diodes (HPDs) [10-14]. Specifically, the combination of disparate materials like glass, metals, ceramics, semiconductors, or MCPs, necessitates a slow, batch-mode process that is incompatible with modern production methods. Optimal vacuum-processing conditions for each of these materials are vastly different. Thus, when disparate materials are processed together (like in the PMT production), the most sensitive components set the upper limit on processing conditions (like temperature, temperature variations, and gradients). These sensitive parts include glass components, glass-to-metal seals, photodiodes, MCPs, and graded-glass electrode feedthroughs. This limitation in the processing temperature and speed inevitably leads to long, static processing.

Our production technology [1] has bypassed this critical manufacturing problem through the highly non-trivial avoidance of all materials other than glass [1, 2, 15] (or a similar dielectric material) among the components entering the uninterrupted vacuum production line (Fig. 1). Our photosensors use vacuum-sealing thin films [16], simultaneously functioning as passes for the high voltage and the ground potential [1]. This is a patented breakthrough concept that, to the best of our knowledge, has not been used in any other technology. As a result, our photosensors can function without any through-the-glass feedthroughs, i.e., without any 'solid' metals (wires, sheets, foils) among the vacuum-processed components (Figs. 1, 2).

## 3. The Operation Principle of the Individual ABALONE Photosensors

Since the performance of individual ABALONE photosensors was already discussed in Refs. [2-4], we will only briefly review their operation principle. Each photoelectron emerging from the photocathode of an ABALONE Photosensor on the inside surface of the Dome (Fig. 2) is accelerated into the small hole in the center of the Base Plate, which is vacuum-sealed by the Windowlet from the outside. The Windowlet-scintillator converts the kinetic energy of an accelerated photoelectron into secondary photons, proportionally to the kinetic energy and the scintillator yield. An accelerated photoelectron can generate hundreds of secondary photons in the LYSO scintillator, and significant fractions of them are detected by a G-APD ('SIPM') outside the vacuum enclosure [2-4].

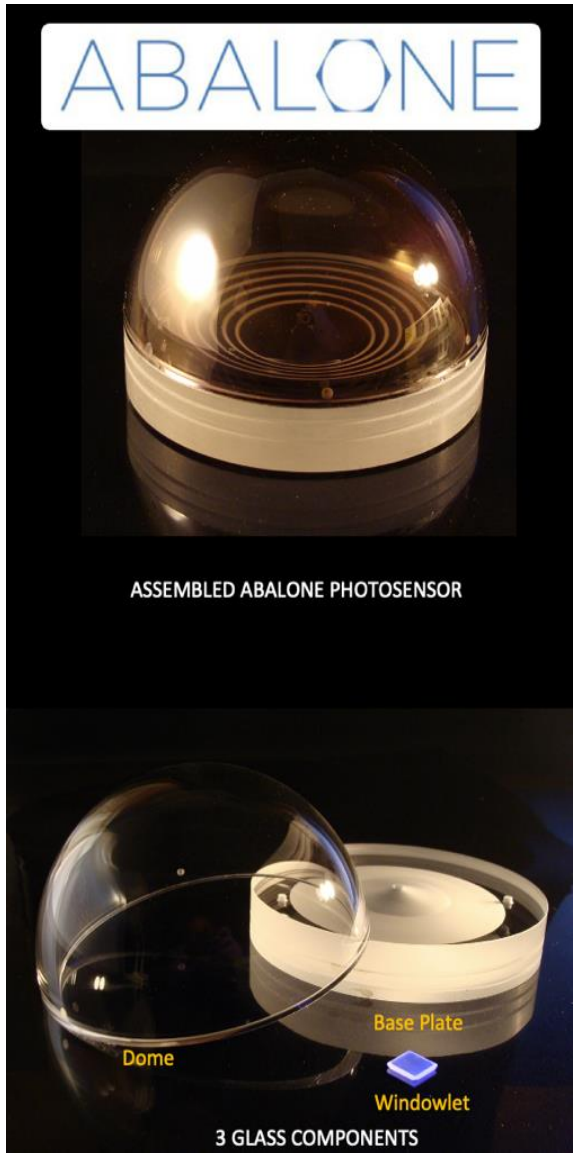

**Fig. 1** ABALONE[TM] photosensor (top) and its only three components (bottom): Dome, Base Plate, and Windowlet. These components form a vacuum-sealed electron lens in conjunction with vacuum-deposited thin films of ultrapure metals. The inner hemispherical Dome diameter is 10 cm, and the Base Plate and the Dome are both made of fused silica. Note the absence of electrodes and feedthroughs.

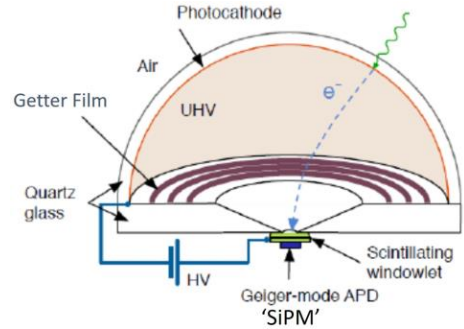

**Fig. 2** Schematic view of a single ABALONE Photosensor, from Ref. [2]. See text for explanation.

## 4. The Tandem-ABALONE[TM] Detector Module

A Tandem-ABALONE Detector Module combines two back-to-back oriented ABALONE photosensors, Fig. 3. We found that an optimal electron lens (Fig. 4) can be designed with the help of the Field Interface – a thin standoff insert between the two ABALONE units, composed of dielectric material and a set of conductive concentric rings. Besides properly shaping the electric field between the two ABALONE units, the Field Interface also connects these units, holds in place the two Light Guides that bring scintillation photons from Windowlets to the G-APDs ('SiPMs'), interfaces a link to a high-voltage power supply, or it can host a miniature power supply within itself.

The Tandem-ABALONE[TM] Detector Module presented in this article (Fig. 3) was specifically developed in collaboration with experts for new detection technology in the IceCube Neutrino Observatory. The requirements dictated by this demanding application [17] are in large part satisfied by the performance of the individual ABALONE Photosensors, outlined in the Introduction [2-4]. What is needed in addition is electrical neutrality of the entire outside surface of the Detector Module, low-temperature operation, small cross-sectional diameter (for the reduction of the diameter of the holes melted in the South Pole ice), and crucially - the $2 \times 2\pi$ angular acceptance (with separate readout for the lower and the upper hemisphere), in order to provide acceptance for downward approaching neutrinos from the Galactic plane [18], and also to more efficiently suppress the background.

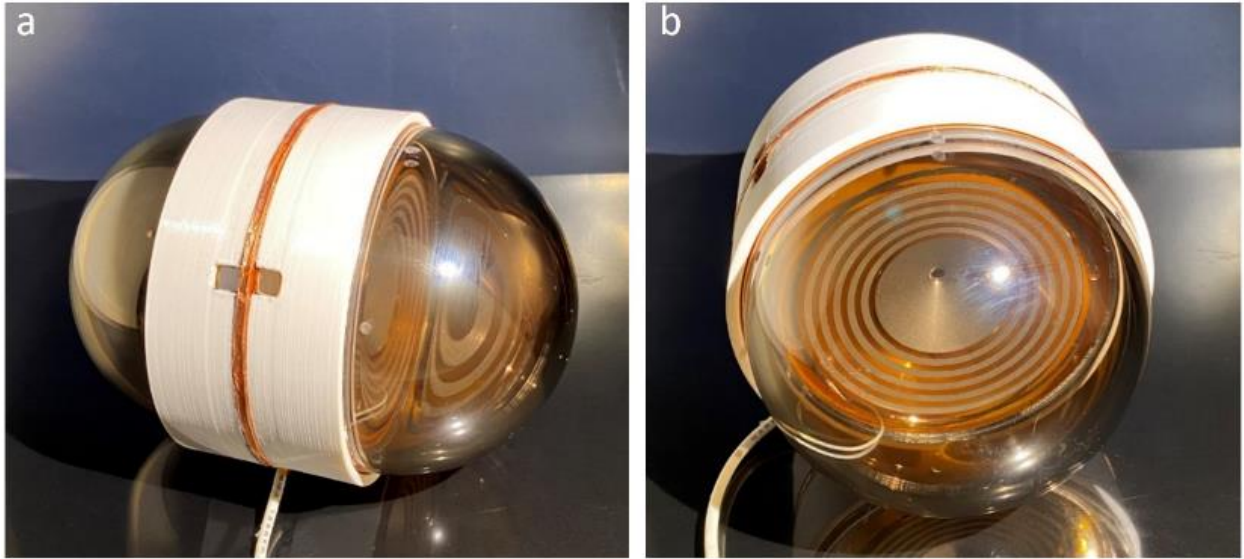

**Fig. 3** Tandem-ABALONE™ Detector Module. (a) The output surfaces of the two rectangular Light Guides are seen in the Field Interface. (b) The small dark circular area in the middle of the ABALONE unit is the active surface of the Windowlet-scintillator, into which photoelectrons are focused and accelerated from the entire photocathode, provided that the Tandem Electron Lens is correctly designed and manufactured.

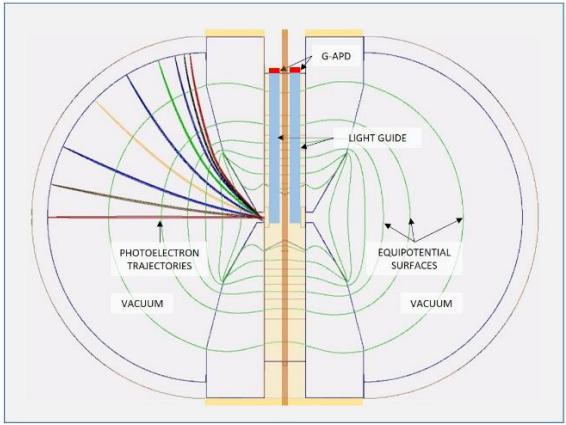

**Fig. 4** Schematic view of a Tandem Module. ABALONE photosensors and the Field Interface (yellow) between them form the electron lens. The design of the Tandem Electron Lens is based on numerical simulations and feedback from tests. Light Guides (blue) transmit the scintillation light from Windowlets to the corresponding G-APDs ('SiPMs') (red). The outside surface of the Module is electrically grounded, as well as the cavity that surrounds the G-APDs ('SiPMs'). (Note that the Light Guides in the tested Tandem Module (Fig. 3) were made longer than in the design presented here, in order to make them easily accessible during testing).

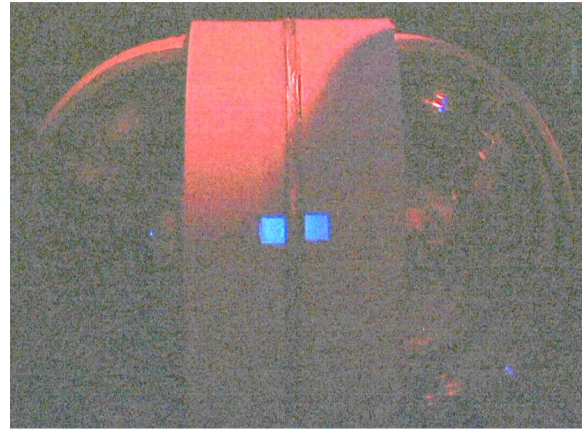

**Fig. 5** Tandem Module in the dark room, illuminated from the above by a weak, non-uniformly dispersed blue light. Scintillation light that is generated by focused 10 keV photoelectrons in the Windowlets of both units is seen in the form of bright blue squares in the Light Guide outputs. The presented long-exposure photo was taken by a CCD camera, while during normal operation the light-guide output is optically coupled to a photodiode.

## 5. Experimental

Both ABALONE units in the tested Tandem Module are made of fused silica. They host a Sb-Cs photocathode and Windowlets made of LYSO scintillator. The outside diameter of the Module is 110 mm.

The evaluation of the Tandem Electron Lens was carried out in a dark room, where the tested Tandem Module was fixed horizontally on top of an optical desk and exposed to two different light sources. One of them was a weak, dispersed blue ambient light (Fig. 5), and another one an attenuated 405 nm laser beam with a widened spot size of 5 mm in diameter (Fig. 6). The laser beam was orthogonal to Module's axis in order to minimize the projected beam spot size in the most critical region for electron optics - the periphery of the hemispherical photocathode. We applied a thin, diffusive Teflon ribbon to the Dome to avoid 'hot' beam spots in the photocathode. The light beam was steered by a remotely operated stepper motor. Red darkroom ambient light was occasionally used to monitor the setup by a CCD camera.

Measurements of the intensity of the scintillation light as a function of the beam position on the ABALONE hemispheres were performed using an Ocean Optics STS UV spectrometer (190 to 650 nm) with internal photodiode readout. At the same time, this versatile instrument has allowed us to verify the spectral shape of scintillation light after its passage through the Light Guides and the optical glue.

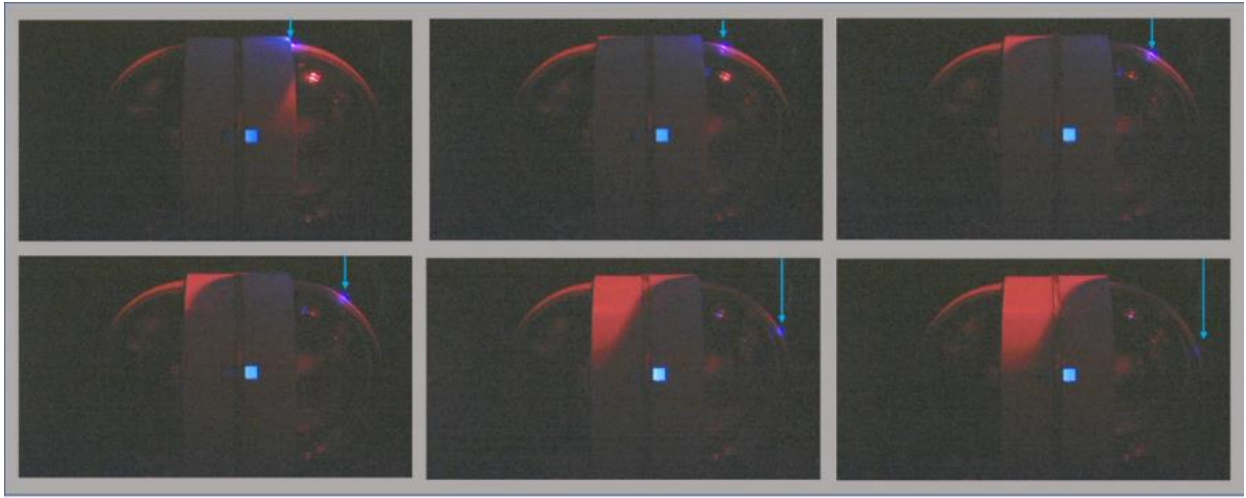

**Fig. 6** Response of the Tandem Detector Module to the light beam, scanned along the Dome of the right ABALONE unit, as indicated by blue arrows. The acceleration potential was 10 kV. (The shifting light-beam source apparatus above the Module cast the changing shadows of red darkroom light).

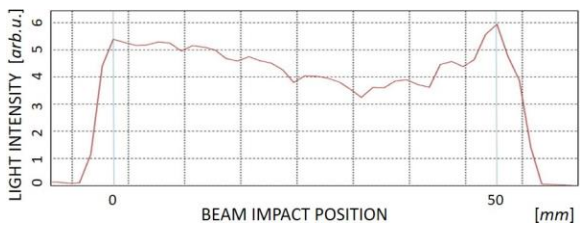

**Fig. 7** Intensity of scintillation light as a function of the test beam position. See text for explanation. (Due to the finite beam size, the intensity does not abruptly vanish for the beam positions outside the photocathode).

CCD images of the uncovered exit surfaces of the two Light Guides (Fig. 6) provide an independent visual verification of the performance of the Tandem Electron Lens.

## 6. Results

In Figs. 6 and 7, the scintillation light of approximately the same intensity is evident for all beam impact locations on the hemispherical photocathode, including its critical periphery. This result confirms the functionality of the Tandem Electron Lens design shown in Fig. 4.

In general, variations in detection efficiency of a photocathode-based photosensor regularly arise from a number of complex position- and angle-dependent optical effects [19]. Systematic variations seen in Fig. 7 are mainly due to reflections off various surfaces, including the diffusive white edge of the Field Interface, the curved surface of the Dome, and entrapment of photons in the glass window that provides multiple conversion opportunities.

We carried out the presented experiments at a fixed photoelectron acceleration potential of 10 kV. For completeness, we present the dependence of the scintillation light intensity on the acceleration potential in Figs. 8 and 9. Our measurement covers the range up to 15 kV, while complementary measurements with higher potentials can be found in Ref. [3].

The nonlinearity seen at the lowest potentials is due to the energy loss of photoelectrons in the thin-film coating of the Windowlet-scintillator surface exposed to the vacuum. This coating holds the anode electric potential on the scintillator, and simultaneously prevents feedback of scintillation light from the Windowlet to the photocathode. Specifically, photoelectrons of *2 keV* lose (on average) the total energy of *2 keV* in this film before they can reach the scintillator. However, electrons of higher energies lose proportionally lower amounts of energy, consistently with Bethe's energy loss relation [2, 20]. For instance, a *20 keV* electron loses on average *~0.2 keV*. The linearity of the studied correlation thus evidently improves progressively with the increasing energy (Fig. 9), and further beyond our current limit [3].

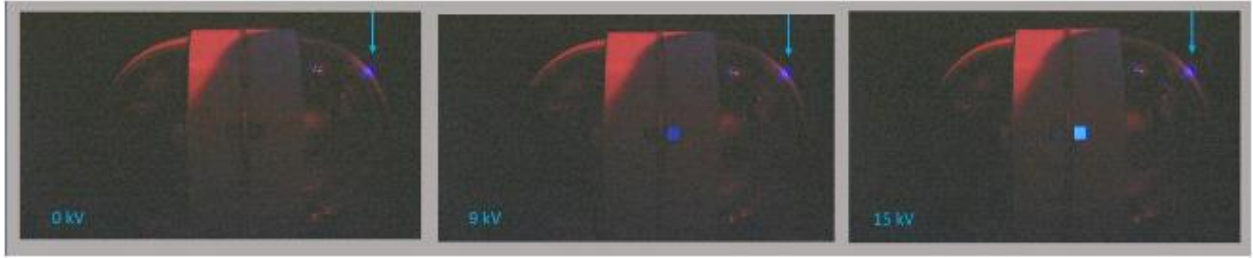

**Fig. 8** Response of the Tandem Detector Module to the acceleration potentials of 0, 9 kV and 15 kV, for fixed light-beam position indicated by blue arrows

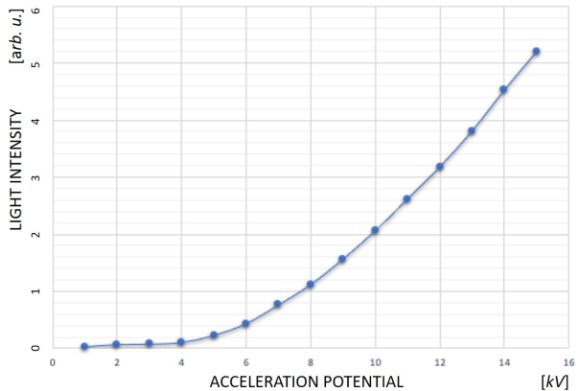

**Fig. 9** The Intensity of the scintillation light as a function of the acceleration potential, for a light-beam position fixed as in Fig. 8.

### 7. Conclusion

We have designed, manufactured, and successfully tested fully functional, product-level Tandem Detector Modules. We discuss key design aspects of these novel photosensors, and report test results. Building upon the powerful foundational ABALONE technology, the Tandem concept provides a unique solution for neutrino astronomy and other large-area radiation detectors. The spherical Tandem Detectors host two back-to-back oriented ABALONE photosensors in unprecedentedly compact and robust Modules, which provide $2 \times 2\pi$ angular acceptance, unparalleled performance, and the means for modern, cost-effective, high-volume production.

Following our success with the development of the Tandem Module technology, our collaborators in the field of neutrino astronomy and dark matter research requested an even more versatile photosensor that would provide internal position resolution, in addition to all of the benefits provided by the Tandem Modules. PhotonLab has recently responded by a new invention called ALKA[TM], whose development started with the support of the U.S. Department of Energy[1].

**Acknowledgments**

PhotonLab, Inc. acknowledges Phase-I and Phase-II SBIR Awards No. IIP-1927067 by National Science Foundation, for the project titled "Novel Photosensor Technology - Enabling the IceCube Cosmic-Neutrino Experiment (South Pole) to open a New Window on the Universe" (PI Ivan Ferenc Šegedin).

We acknowledge invaluable contributions to the presented development by Prof. Sebastian Böser of UNI-Mainz and his team from the IceCube Collaboration, Prof. Alfredo Davide Ferella (University of L'Aquila), Prof. Jan Conrad and Dr. Jörn Mahlstedt (Stockholm University), and their teams in experiments XENON and DARWIN.

---

[1] SBIR Award Number DE-SC0022519, "Novel Position Sensitive Photosensor Technology for Rare Decay and Rare Particle Detection"